\documentclass[prl,twocolumn,nofootinbib,preprintnumbers,amssymb,amsfonts,amsmath,superscriptaddress,showpacs,hyperref]{revtex4-1}

\usepackage{graphicx}
\usepackage{bm}
\usepackage{color}
\usepackage{url}
\usepackage{epstopdf}
\usepackage{amsthm,mathrsfs} 

\newcommand{\Mpl}{M_{\rm Pl}}
\newcommand{\Sc}{\mathcal{S}}
\newcommand{\xo}{x_{{\rm zpm}}}

\begin{document}

\title{Tests of Quantum Gravity induced non-locality\\ via opto-mechanical quantum oscillators}

\author{Alessio Belenchia} 
\email{abelen@sissa.it}

\author{Dionigi M. T. Benincasa}
\email{dionigi.benincasa@sissa.it}

\author{Stefano Liberati}
\email{liberati@sissa.it}
\affiliation{SISSA, Via Bonomea 265, I-34136 Trieste, Italy and INFN, Sez. di Trieste.}

\author{Francesco Marin}
\email{marin@fi.infn.it}
\affiliation{Dipartimento di Fisica e Astronomia, University of Florence, Via Sansone 1, I-50019 Sesto Fiorentino, Firenze, Italy and INFN, Sez. di Firenze.}
\affiliation{European Laboratory for Non-Linear Spectroscopy (LENS), Via Carrara 1, I-50019 Sesto Fiorentino, Firenze, Italy.}

\author{Francesco Marino}
\email{marino@fi.infn.it}
\affiliation{CNR-Istituto Nazionale di Ottica, Largo E. Fermi 6, I-50125 Firenze, Italy and INFN, Sez. di Firenze.}

\author{Antonello Ortolan}
\email{antonello.ortolan@lnl.infn.it}
\affiliation{INFN, Laboratori Nazionali di Legnaro, Viale dell'Universit\`a, 2, 35020 Legnaro, Padova, Italy.}


\begin{abstract}
Several quantum gravity scenarios lead to physics below the Planck scale characterised by nonlocal, Lorentz invariant equations of motion.  We show that such non-local effective field theories lead to a modified Schr\"odinger evolution in the nonrelativistic limit. In particular, the nonlocal evolution of opto-mechanical quantum oscillators is characterised by a spontaneous periodic squeezing that cannot be generated by environmental effects. We discuss constraints on the nonlocality obtained by past experiments, and show how future experiments (already under construction) will either see such effects or otherwise cast severe bounds on the non-locality scale (well beyond the current limits set by the Large Hadron Collider). This paves the way for table top, high precision experiments on massive quantum objects as a promising new avenue for testing some quantum gravity phenomenology.
\end{abstract}

\maketitle

{\em Introduction.}---A vast uncharted territory lies between the scale presently tested in high energy accelerators, $\approx 10^{13}$ eV, and the Planck scale, $\Mpl=1.22\times 10^{28}$ eV, where quantum gravitational physics is expected to become relevant.
Quantum gravity phenomenology is a broad field of research which attempts to bridge this gap by connecting models of quantum gravity with observation/experiments. In the absence of a definitive theory of quantum gravity, much of the present literature has resorted to testing general ideas ensuing from various models, ranging from large extra dimensions~\cite{Hossenfelder:2003jz}, to generalisations of the uncertainty principle at the Planck scale~\cite{Garay:1994en, Hossenfelder:2012jw}, to effective field theories (EFT) implementing a breakdown of Lorentz invariance at the Planck scale~\cite{Mattingly:2005re,Liberati:2013xla}, and several others (see also~\cite{AmelinoCamelia:2008qg} for a more comprehensive review of current proposals). 

Within this context, special interest is deserved by those QG scenarios where Lorentz invariance is held as a guiding principle, 
while spacetime is seen as emerging from more fundamental
structures \cite{Bombelli:1987aa}.  
It is generally believed that a successful marriage between fundamental discreteness and 
local Lorentz invariance (LLI) requires some form
of non-locality~\cite{Benincasa:2010ac,Gambini:2014kba}. 
Indeed, disparate approaches to quantum gravity where this marriage occurs explicitly lead to 
nonlocal modifications of standard local dynamics. Explicit realisations arising from quantum gravity models aside, the appearance of nonlocal 
dynamics can be expected on very general grounds: LLI together with the the requirement that
the dynamics not suffer from classical instabilities, effectively singles out two types of dynamics, standard local
dynamics (1st or 2nd order in space and time) and nonlocal dynamics ($\infty$-order in space and time)
--- the  infinite number of time derivatives being needed in order to avoid Ostrogardsky's
theorem~\cite{Ostro}.

In particular, 
%
%
%
%
%
let us consider the example of a free scalar field in flat spacetime. 
Its standard dynamics are given by the 
Klein-Gordon (KG) equation,
i.e. the dynamics
are {\em local}, and indeed {\em stable}. A modification of this dynamical law, which respects 
LLI and avoids generic Ostrogardsky instabilities, must necessarily take the form 
$(\Box+m^2)\to f(\Box+m^2)$, where $f$ 
is some non-polynomial function, i.e. the dynamics become {\em nonlocal}, 
with stability depending on the specific choice
of $f$. It follows therefore that if specific models of QG
lead to modifications of standard local dynamics while simultaneously claiming to preserve
LLI, then they must do so through nonlocal dynamics of the kind just described. Indeed, this
expectation is realised in various QG models which hold LLI as a guiding principle
\cite{Biswas:2014yia,Koshelev:2011gx,Sorkin:2007qi}. 

%
%

The definition of $f$ must contain a characteristic, covariantly defined scale, 
which allows for a suitable power law expansion characterising the deviation 
from the standard local field equations. It is important to note that this scale,
which we refer to as the non-locality length scale, $l_k$, need not be related
to the quantum/discreteness scale normally associated to QG, i.e. the Planck scale.
In fact, in various approaches to QG leading to this kind
of nonlocal dynamics this will usually be some mesoscopic length scale lying 
somewhere between the TeV scale 
and the Planck scale (see e.g. \cite{Sorkin:2007qi}); a fact which is of particular
relevance within the context of casting phenomenological constraints.

In what follows we shall describe a novel, promising way to test non-local dynamics of this type.
Restricting ones attention to nonlocal effective field theories which are both
classically and quantum mechanically stable and, crucially, that can be derived from an underlying
quantum gravity theory, effectively singles out nonlocal dynamics defined by an analytic
$f$. This does not imply that non-analytic $f$s are not interesting
(in fact this case naturally arises in the context of causal set theory \cite{Aslanbeigi:2014zva,Belenchia:2014fda}), 
but rather that further investigations in 
the stability and unitarity of these theories are needed before much effort is invested  
in extending an analysis similar to the one described below
to such cases. A significant consequnce of considering analytic $f$s is that it will allow us to solve the dynamics 
perturbatively.

We will perform a perturbative study of the effects of modified equations of motion on the evolution of opto-mechanical 
quantum oscillators based on the following methodological steps. 
In order to compare our analysis to actual experiments on the aforementioned systems,
we must first derive the non-relativistic limit of an 
effective nonlocal massive scalar field theory with dynamics given by an analytic $f$, showing
that the evolution of a quantum system is governed by a modified (nonlocal) Schr\"odinger equation. 
This approach allows us to overcome long standing issues about how to relate the evolution of macroscopic 
objects to the effects of nonlocality arising from a potential discreteness of spacetime. We  
proceed by performing a perturbative expansion around the local regime adapted to the 
specific experimental setting we are interested in. We then solve for the evolution of the wave function and compute the 
behaviour of the relevant physical observables. We show that a characteristic signature due to a periodic squeezing is 
introduced by the first nonlocal correction. Using this feature we discuss the constraints already available and provide 
forecasts for those deducible from future experiments. 
In particular, we establish that sensitivity close to the Planck scale can be achieved, thus severely
constraining, or even ruling out, models of QG in which the nonlocality scale is larger than the Planck scale.


{\em Framework}.--- For the non-relativistic limit of an equation of the form $f(\Box+m^2)\phi=0$ to be physically meaningful and have the usual  probabilistic interpretation of the wave function, we require that: 
a) $f(-k^2+m^2)=0$ iff $k^2=m^2$ (this property ensures that there exist no classical runaway solutions 
and, when $f$ is entire, no ghosts).
b) unitarity of the full nonlocal quantum field theory (QFT)
and
c) the non-local QFT must possess a global $U(1)$ symmetry with a conserved charge
which is positive semi-definite in the non-relativistic limit
(this condition ensures that a probabilistic interpretation can be given). 

For concreteness let us consider now the non-local Lagrangian for a free complex, massive, scalar field
\begin{equation}
\mathcal{L} = \phi(x)^*f(\Box+\mu^2)\phi(x)+c.c.,
\label{lag}
\end{equation}
where $\Box = c^{-2}\partial^2_t -\nabla^2$, $\mu=mc/\hbar$ and 
we assume that $f$ is an analytic function so that it can be formally expanded as a power series 
$f(z) = \sum_{n=1}^\infty a_n z^n$. 
Implicit in the definition of $f$ is $l_k$, which in the local limit $l_k\rightarrow0$ sends $f(\Box+\mu^2)\rightarrow \Box+\mu^2$; in particular 
$a_n\propto l_k^{2(n-1)}$.

Upon making the ansatz  $\phi(x)=e^{-i \frac{m c^2}{\hbar}t}\psi(t,x)$ and taking the limit $c\rightarrow\infty$ we find
\begin{equation}
\mathcal{L}_{\text{NR}}  = \psi^*(t,x)f(\Sc')\psi(t,x)+c.c.,
\label{lag2}
\end{equation}
where NR stands for non-relativistic, $\Sc' =-\frac{2m}{\hbar^2}\Sc$ and
\begin{equation}
\Sc=i\hbar \frac{\partial}{\partial t} + \frac{\hbar^2}{2m}\nabla^2,
\end{equation}
is the usual Schr\"odinger operator.

Depending on the precise form of the function $f$, the conserved charge associated
to the symmetry $\psi\rightarrow e^{i\alpha}\psi$ may or may not be positive semidefinite.
As stated in (c), in what follows we will assume that such a condition is satisfied. In any 
case, one can show~\cite{NLS}  that the conserved current is given by a perturbative expansion
in $\epsilon$ whose zeroth order term is the usual one.

{\em Perturbative expansion.}--- We shall now consider the case of 
a harmonic oscillator in 1-dimension whose evolution is assumed to be described by the above non-local Schr\"odinger equation with a 
harmonic potential. This study is motivated both by its simplicity and ubiquity in physics, and in view of its application to the actual experiments involving  
systems that are effectively 1-dimensional. 

Hence, we wish to solve the nonlocal equation 
\begin{equation}
\label{nlse}
f(\Sc)\psi(t,x) = V(x)\psi(t,x),
\end{equation}
where $V$ is a harmonic potential $V(x)= \frac{1}{2}m\omega^2 x^2$, $m$ is the mass of the system and $\omega$ its
natural angular frequency and, consistently with the local limit requirement, we write  
\begin{equation}
f(\mathcal{S})= \Sc+\sum_{n=2}^{\infty}b_{n}\left(\frac{-2m}{\hbar^{2}}\right)^{n-1}l_k^{2n-2}\Sc^{n},
\end{equation}
where the $b_{n}$ are dimensionless coefficients.  

The introduction of the potential allows one to construct a dimensionless parameter
$\epsilon\equiv m\omega l_k^2/\hbar$ 
which can be used to define the perturbative expansion. 
Note that $\sqrt{\epsilon}$ represents the ratio between $l_k$ and the 
width of the oscillator's ground-state 
wavefunction $\xo = \sqrt{\hbar/m\omega}$ and, 
through $\epsilon$, the 
mass parameter enters the nonlocal dynamics of the harmonic oscillator 
breaking the usual $\omega$ scaling. This dependence on $m$ suggests that 
massive quantum systems could be the ideal setting for 
detecting such non-locality.

In order to keep the notation as clear as possible we will use dimensionless variables\footnote{These are defined as $\hat{t}\equiv \omega t,\ \hat{x}=x/\xo$. In the main text we have systematically  omitted the hats for notational convenience.} for which we have
\begin{equation}
\left({\mathcal{S}}+\sum_{n=2}^{\infty}b_{n}\epsilon^{n-1}(-2)^{n-1}{\mathcal{S}}^{n}\right){\psi}=\frac{1}{2}{x}^{2}{\psi}.
\label{schpert}
\end{equation}
Given the complexity of such an integro-differential equation, together with the fact that we are only 
interested in small deviations from the standard local behaviour, 
we shall assume that 
the non-local Schr\"odinger equation admits solutions of the form 
\begin{equation}
\psi= \sum_{n=0}^{\infty}\epsilon^{n}\psi_{n},
\label{expansion}
\end{equation}
where $\psi_{0}$ is a solution to the standard Schr\"odinger equation with harmonic potential,
and higher order terms are suppressed by powers $\epsilon$. We are therefore 
assuming that the nonlocal equation admits $\psi_0$ as an approximate solution in the 
sense that $(f(\mathcal{S})-V)\psi_0=O(\epsilon)$ (see \cite{NLS} for further details).
This assumption is based on the intuition that a physically
reasonable nonlocal generalisation of the Schr\"odinger equation
should admit solutions which are perturbations of solutions of the
local Schr\"odinger equation. The failure of this property would otherwise 
put the theory in serious tension with well-tested physics. 
%

We will be interested in perturbations around coherent states, since these are easiest to 
realise within the experimental settings we have in mind, 
and furthermore include the harmonic oscillator's ground state as a  specific case. 
Substituting this back into \eqref{schpert} one finds that the differential equation 
at order $\epsilon^n$ is the standard Schr\"odinger equation for $\psi_n$ with a source term 
depending on the solution to the perturbative problem at order $\epsilon^{n-1}$.

At zeroth-order in $\epsilon$ we have the standard Schr\"odinger equation with a harmonic oscillator 
potential for which we
choose a solution corresponding to a coherent state, $\psi_{0}^\alpha$.
To solve the equation to first order we make the following ansatz
\begin{eqnarray}
\psi_{1}(t,x)= \psi_{0}^{\alpha}(t,x)\left[c_{0}(t)+c_{1}(t) x+c_{2}(t)x^2 \right. \nonumber\\
\left.+c_{3}(t)x^3 +c_{4}(t)x^4 \right].
\end{eqnarray}
Substituting this into equation \eqref{schpert} and keeping only terms of order $\epsilon$  
one finds a system of ODEs for the time dependent coefficients that can be easily solved.

{\em Results.}--- We present results for perturbations around a generic coherent state 
obtained by solving the non-local Schr\"odinger equation perturbatively 
up to order $\epsilon$
(the corresponding results for the ground state 
can be found by setting the coherent amplitude $\alpha=0$). 
Since $\langle\psi|\psi\rangle$ is not conserved by the first perturbative term, in order to have a well-defined 
probability distribution we normalise 
the wavefunction using the norm of $\psi_{0}^{\alpha}+\epsilon \psi_{1}$, 
i.e.~in accordance with the Born rule we shall define the probability density as
\begin{equation}
\rho(t,x)= \frac{\psi^*(t,x)\psi(t,x) }{\int^\infty_{-\infty} |\psi|^2 {\rm d}x},
\end{equation}
such that $\int_{-\infty}^\infty dx\,\rho(x)=1$.
While for the ground state this normalization factor is one at order $\epsilon$, in the case of a generic coherent state 
an order $\epsilon$ time dependent correction will be present. The above normalisation factor ensures 
that even in this case we a have a meaningful probability distribution.

\begin{figure}[tbp]
\begin{center}
\includegraphics[width=0.45\textwidth]{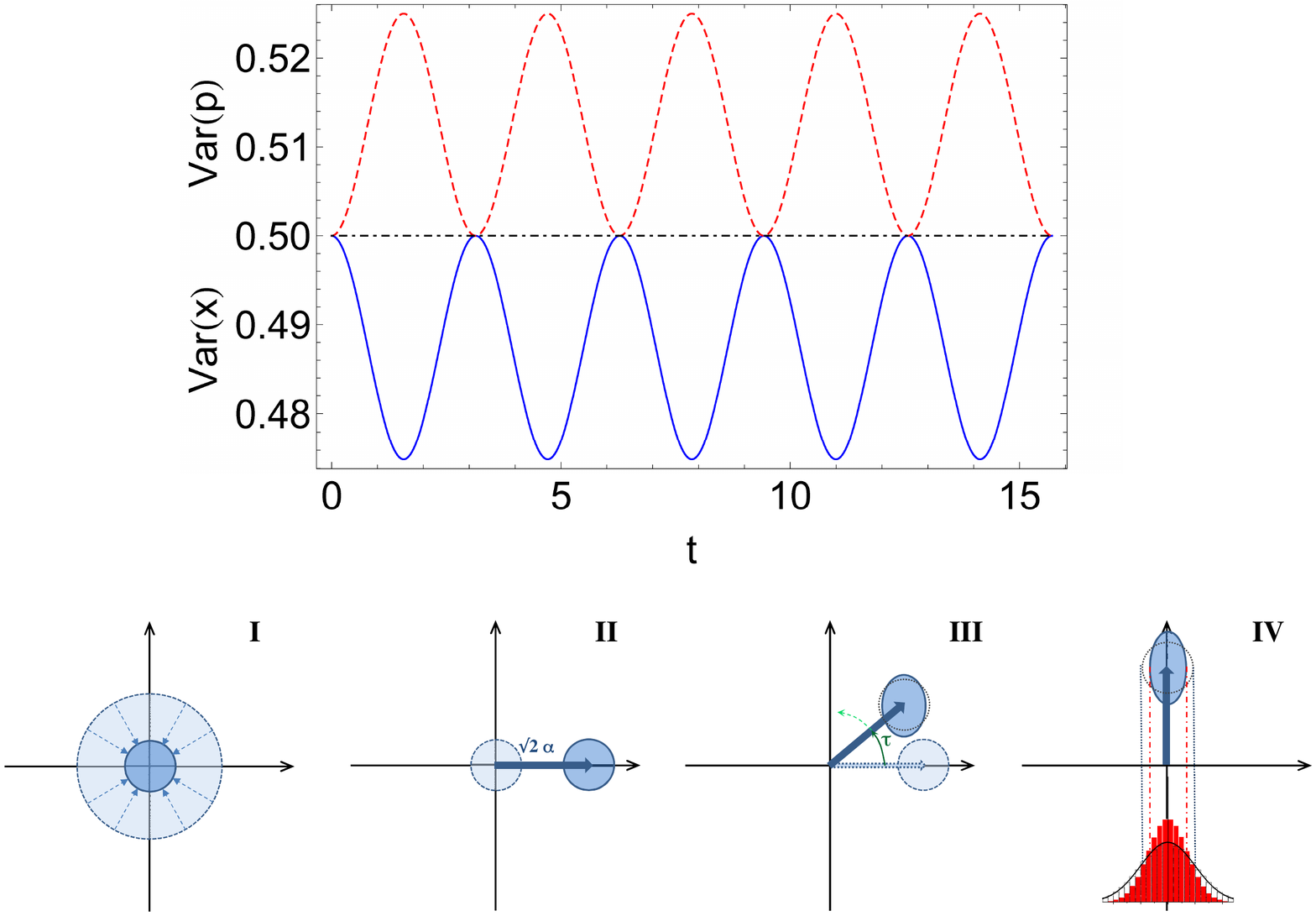}
\caption{\small{Time dependence of the variances of a coherent state for $ \alpha=1$ and $b_2 \epsilon=10^{-2}$.
The continuous (blue) and dashed (red) lines represent the position and momentum variances respectively. 
The black dot-dashed line is the standard value, 1/2, of both variances in the local theory.  Below the plot we sketch in 
$x-p$ phase diagrams the proposed experimental procedure for measuring the variance of $x$, involving (I) cooling 
the oscillator down to $\langle n \rangle \ll 1$, (II) a pulsed excitation in a well-defined coherent state, (III) free evolution 
for a time $\tau$, (IV) the measurement of $x$ in a time interval shorter than the oscillation period. 
Steps (II) and (III) should last much less than the thermal decoherence time.  
The cycle is iterated several times, the variance of the measurements of $x$ is calculated, then $\tau$ is changed and 
the whole measurement  procedure repeated.}} 
\label{fig:variances}
\end{center}
\end{figure}

The mean of position and momentum are given by
{\small
\begin{subequations}
\begin{align}
\langle x\rangle &=\sqrt{2} \alpha  \cos (t)\, \left\{1+ \frac{ 1}{4} \epsilon  \alpha ^2 b_{2} \left[ \cos (2 t)-1 \right] \right\}+\mathcal{O}(\epsilon^{2}),
\label{meanxcoherent}\\
\langle p\rangle &= \sqrt{2} \alpha  \sin (t)\left\{ 1+ \frac{1}{4}   \epsilon b_2  \left[ \alpha ^2 (7+3  \cos (2 t))-2 \right]\right\} +\mathcal{O}(\epsilon^{2}).
\label{meanpcoherent}
\end{align}
\end{subequations}}
At order $\epsilon$ the variance in position and momentum are given by
\begin{align}\label{varx}
\mbox{Var}(x)&=\frac{1}{2}\left\{ 1- \epsilon  b_2 \left[\left(6 \alpha ^2-1\right) \sin^2 ( t) \right]\right\}+\mathcal{O}(\epsilon^{2})\, ,\\
\mbox{Var}(p)&=\frac{1}{2}\left\{ 1+  \epsilon  b_2 \left[\left(6 \alpha ^2-1\right) \sin^2(t) \right]\right\}+\mathcal{O}(\epsilon^{2})\, .
\label{varp}
\end{align}
Note that $\mbox{Var}(x)\mbox{Var}(p)=1/4+O(\epsilon^{2})$, which is consistent with the 
state being a state of minimum uncertainty to first order, although the variances themselves possess a 
time dependence which leads to
a spontaneous time-dependent, cyclic squeezing in both position and momentum, see Fig.~\ref{fig:variances}.

%


{\em Present constraints and forecasts.}--- We now consider the constraints imposed by both existing opto-mechanical experiments and experiments that will be performed in the near future. Let us begin by noting that 
bounds on the non-locality scale have already been obtained by comparing nonlocal relativistic EFTs
to the 8 TeV LHC data~\cite{Biswas:2014yia}, in which the authors find $l_{k}\leq 10^{-19}$m.

Returning to the non-relativistic setting, experiments have recently achieved the goal
of cooling a mechanical oscillator down to thermal occupation numbers below unity, 
i.e.~to conditions close to the ground state. In such systems, a wave function can be associated to an effective coordinate 
describing an elastic vibration mode, or, to reasonable approximation, the center-of-mass of an oscillating portion of a body. The coordinate is measured by a mode of an electromagnetic field (the \emph{meter}), whose eigenfrequency depends on such coordinate. The interaction can be generally described by a Hamiltonian of the kind $-\hbar g a^{\dagger}a x$, where $g$ is the coupling strength that depends on the system's 
geometry and $a$ is the electromagnetic field's annihilation operator~\cite{Aspelmeyer2014}.
 
The expressions that we derived contain 
coefficients, $b_n$, that depend on the specific model. 
For the sake of clarity, in the following we set  $b_2 = 1$. 
We start our analysis by  comparing the measured variance of $x$ with our corresponding prediction for the ground state.
Taking the time average of~\eqref{varx}, with $\alpha = 0$, we infer that it should differ from its standard value by $\epsilon/2$. For a meaningful comparison, the experimental system needs to be well within the 
perturbative regime, i.e.~$\epsilon \ll 1$. This implies that {the experiment should achieve} an oscillator energy close 
to its standard ground value (namely with an average occupation number $\langle n\rangle  \ll {1}$). 
In these conditions, Eq.~\eqref{varx} allows one 
to derive an upper limit to $\epsilon$ of 
the form $\epsilon < 2 (2\langle x^2\rangle _{\rm measured}-1)$.

The experiments with macroscopic oscillators
that enter the quantum regime more deeply
(Ref.~\cite{OConnell2010} achieved $\langle n\rangle  <0.07$ and~Ref.~\cite{Meenehan2015} achieved $\langle n\rangle =0.02$), 
do not provide a measurement of the variance of $x$. 
Other experiments~\cite{teufel2011sideband,wollman2015quantum,Pirkkalainen2015,Lecocq2015}, which use the drum mode of aluminum membranes, 
measure instead $\langle x^2\rangle $ of
the oscillator coupled to microwave radiation that is used to cool and monitor its motion.
In the best case scenario, with $\langle n\rangle  < 0.1$~\cite{Lecocq2015}, Eq.~\eqref{varx} yields the upper limit $\epsilon < 0.2$. 
Using $\xo \simeq 5$~fm (the mass is $\sim 50$~pg~\cite{teufel2011sideband} and the frequency $\omega/2 \pi = 15$~MHz), we derive $l_k < 2\times10^{-15} $~m. This is reasonably close to the constraint obtained at LHC and has the extra virtue of being 
independent of it.
      
A further comparison between our theory and experiments can be based on the evolution of coherent states. As shown in Eqs.~\eqref{meanxcoherent}-\eqref{meanpcoherent}, our model predicts some third-harmonic distortion proportional to $\epsilon$ and $\alpha^3$: the effect of non-locality is thus amplified, with respect to the ground state, by the coherent amplitude $\alpha$. Unfortunately, to our knowledge no experiment could so far realise a quantum coherent state, i.e.~a coherent state produced from the ground state, where thermal fluctuations are negligible. Treating such cases rigorously would require a generalisation of Eq.~\eqref{meanxcoherent}-\eqref{meanpcoherent} to thermal states, something far from being straightforward. 

However, the situation is not as hopeless as it might seem. A recent set of experiments developed highly 
isolated quantum oscillators, i.e.~with a high mechanical quality factor, to constrain potential 
Planck scale deviations from the standard uncertainty relations~\cite{Bawaj:2014cda}. These systems also 
suffer from the issue of only being able to generate thermal coherent states, but are presently undergoing
improvements which should allow them to enter a truly quantum regime. 

Interestingly, we can already provide first estimates of the strength of the constraints achievable via these experiments. In Ref.~\cite{Bawaj:2014cda}, the oscillator is excited in order to create a coherent state with large amplitude $\alpha$, then the time evolving position $x$ is monitored through weak coupling to a \emph{meter} field. The third-harmonic component of $x(t)$ is accurately evaluated as a function of $\alpha$. Such third-harmonic should be null in a perfectly harmonic oscillator, with standard dynamics. On the contrary, its presence is predicted by Eqs.~\eqref{meanxcoherent}, namely with a ratio between third- and first-harmonic amplitudes (third-harmonic distortion) equal to $\epsilon \alpha^2/8$. A third harmonic can also be interpreted as a consequence of the deformed commutator analysed in Ref.~\cite{Bawaj:2014cda}, with a third-harmonic distortion equal to $\alpha^2 \beta/4$ where $\beta$ is a deformation parameter. 

A non-null third-harmonic distortion 
is indeed found in Ref.~\cite{Bawaj:2014cda}, and it is explained as the effect of structural non-linearities. Assuming no accidental cancellation between 
non-linearities and deformed dynamics, and attributing, in the worst case, the overall effect to the deformed dynamics, Ref.~\cite{Bawaj:2014cda} derives a series of upper limits for $\beta$ for different oscillator parameters (frequency and mass). Comparing the above expressions for the third-harmonic distortion, we can directly translate the upper limits on $\beta$ to upper limits on $\epsilon$ ($\beta \to \epsilon/2$) and $l_k$ ($\epsilon \to m\omega l_k^2/\hbar$). Such limits range from $l_k\lesssim 2\times 10^{-22}$ m (for an oscillator with $m = 2\times 10^{-11}$~kg and $\omega = 7.5\times10^5 $~Hz), to $l_k\lesssim1\times 10^{-29}$ m ($m = 3\times 10^{-5} $~kg and $\omega = 5.6\times10^3 $~Hz). 

Such figures suggest that similar experiments, 
once performed on pre-cooled, quantum oscillators --- 
all improvements within technical reach and currently under realisation --- 
could explore the interesting region between the electroweak and 
Planck scales and cast constraints stronger than those achieved at LHC 
by several orders of magnitudes. Let us remark that while our best forecast falls 
short by roughly six orders of magnitude from 
$l_p=10^{-35}$m, 
figures of this order could already be effective for constraining scenarios
where the non-locality scale is larger than the Planck one. 

The experiments just described are useful to set constraints on non-local effects
but, should they be observed, can hardly 
be used to claim the discovery of such phenomena,
since a third-harmonic distortion can be attributed to several effects. 
On the other hand, the spontaneous, oscillating squeezing is a much more meaningful phenomenon 
that cannot be generated by environmental effects. Of crucial importance for a potential experimental test is that 
the oscillation has a precise phase relation with the evolution of the average position 
in a coherent state, a property that can be exploited in synchronous detections. 

A realistic experiment could be based on repeated measurement cycles on a mechanical oscillator, including: a) the cooling down to $\langle n \rangle \ll 1$, b) the pulsed excitation in a well-defined coherent state and 
c) after a delay $\tau$, the measurement of $x$ (see the sketch in Fig.~\ref{fig:variances}).  
Note that the use of pulsed techniques \cite{Vanner2013}, and the estimate of time-evolving indicators 
on iterated measurements \cite{Pontin2015}, are indeed at the forefront of experimental research in opto-mechanics. 
In this sense, we argue that the class of experiments under preparation are already in the 
right 
regime 
to severely constrain (or in the best scenario confirm) non-local physics at sub-Planckian scales.

{\em Conclusions}. ---
In this work we have shown that experiments via opto-mechanical quantum harmonic 
oscillators can cast very severe constraints on the scale of non-locality associated to several 
QG scenarios which respect local Lorentz invariance. 
This was achieved by considering the non-relativistic limit of a non-local EFT 
characterised by an analytic function of the Klein-Gordon operator. 
The so derived non-local Schr\"odinger equation was solved perturbatively in the experimentally relevant case of a harmonic oscillator.
The perturbed solutions showed a characteristic periodic squeezing which cannot be introduced by environmental effects or by dissipative 
behaviour due to the finiteness of the quality factor. This  
provides a clear signature of what one should seek for in this class of experiments, 
and opens up a new channel for quantum gravity phenomenology.
Improvements of the oscillators reported in~\cite{Bawaj:2014cda} are under development with the precise aim to test 
the features presented here. These new experiments will soon reach sensitivities close to the Planck scale, and will 
therefore offer a concrete possibility of observing quantum gravity induced effects, or at least rule out a whole class of candidate theories.

\begin{acknowledgments} 
{\em Acknowledgments}. --- A.B., D.M.T.B and S.L. wish to acknowledge the John Templeton Foundation for the supporting grant \#51876. 
AB also acknowledges the support of STSM Grant from the COST Action MP1006 and would like to thank F.~Dowker and Imperial 
College for interesting discussions and hospitality during 
 early stages of this work. The authors also thank S.~Hossenfelder for useful comments on the manuscript.
\end{acknowledgments} 

\end{document}